# Nearly Free Electron States in Graphane Nanoribbon Superlattice

Qiaohong Liu, Zhenyu Li, and Jinlong Yang

Hefei National Laboratory for Physical Sciences at Microscale, University of Sciences and Technology of China, Hefei 230026, China

Nearly free electron (NFE) state has been widely studied in low dimensional systems. Based on first-principles calculations, we identify two types of NFE states in graphane nanoribbon superlattice, similar to that of graphene nanoribbons and boron nitride nanoribbons. Effect of electron doping on the NFE states in graphane nanoribbon superlattice has been studied, and it is possible to open a vacuum transport channel via electron doping.

**Introduction**

After its successful production [1], two-dimensional (2D) graphene material has been widely studied [2-4]. Cutting graphene into nanoribbons is a natural way to induce a band gap in its electronic structure [5]. Graphene nanoribbon has thus become a subject of interest [6-9]. At the same time, research efforts have been also devoted to functionalization of graphene. Graphane, the hydrogenated graphene, is a typical example [10-13], which is an insulator with a predicted band gap of ~3.5 eV within the density function theory (DFT) [10].

Nearly free electron (NFE) state [14-21] exists widely in low dimensional materials, and it is an attractive candidate for electron transport. Unfortunately, in most cases, NFE states are unoccupied. For this reason, it is very desirable to tune their energy and make them occupied. Various methods have been employed for this purpose, including electron doping, alkali doping, and external electric field modification. [17, 20-22]

Recently, two distinct types of NFE states have been found in graphene nanoribbon superlattice [22]. One (NFE-vacuum) is distributed in the vacuum area between neighboring graphene nanoribbons, while the other (NFE-ribbon) is distributed on the two sides of the ribbon plane. NFE-vaccum can be easily occupied upon electron doping to graphene nanoribbons. These two types of NFE states have also been observed in boron nitride nanoribbon superlattice. However, different with graphene nanoribbon superlattice, the lowest occupied NFE state upon electron doping is mainly NFE-ribbon states mixed with some NFE-vacuum states in boron nitride nanoribbon superlattice.

Electronic structure of graphane is similar to boron nitride nanoribbon with NFE state at the valence band edge [13]. It is interesting to check the behavior of NFE states in graphane nanoribbon superlattice upon electron doping. In this paper, based on plane-wave first-principles calculations, we also identify two types of NFE states in graphane nanoribbon superlattice, and their response to electron doping is discussed. A different behavior compared to both graphene and boron nitride nanoribbon superlattices is predicted.

**Methods and Model**

We performed electronic structure calculations using DFT with the PW91 generalized gradient approximation (GGA) [23] as implemented within the VASP [24,25] package. The cut-off energy of plane wave basis set is set to 400 eV, and the total energy is converged to $10^{-5}$ eV. Graphane nanoribbons are treated using the supercell approach within periodic boundary condition. Vacuum space between adjacent graphene nanoribbons is at least 15 Å. Special k-points for Brillouin zone integrations were sampled using the Monkhorst-Pack scheme [26] with 25 and 19 points along the zigzag and armchair NR direction, respectively, and only one point was used in the other two directions. Doping electrons was compensated by a uniform positive background.

**Results and Discussion**

As shown in Figure 1, N=10 zigzag graphane nanoribbon and N=17 armchair graphane nanoribbon have been studied. Both ribbons have a width about 22 Å. In the ribbon plane, the distance between two ribbons is longer than 10 Å, while the vacuum distance between two ribbons in the direction perpendicular to the ribbon plane is more than 15 Å. Fig. 1(c) and 1(d) show the band structures and the Γ point density profiles of several lowest NFE states for zigzag and armchair nanoribbons, respectively. We can clearly see that there are two types of NFE states similar to graphene nanoribbons.[22] We also named the NFE states mainly distribute on the two sides of the ribbon plane as NFE-ribbon, and NFE states mainly distribute in the vacuum region between nanoribbons as NFE-vacuum.

In both cases studied here, NFE-vacuum states appear at higher energy than NFE-ribbon states, and the bottom of conduction bands is a NFE-ribbon state, similar to boron nitride nanoribbons. In a previous work [22], we have observed that, upon electron doping, NFE-vacuum states in boron nitride nanoribbon superlattice will move downward, but finally mixed with NFE-ribbon states. It means that it is not possible in boron nitride nanoribbon superlattice to obtain occupied NFE-vacuum state via electron doping. However, an occupied NFE-vacuum state is especially interested as a scattering-free transport channel. Will graphane nanoribbon superlattice show similar behavior as boron nitride nanoribbons?

Fig. 2(a) and 2(b) present the band structures and Γ point density profiles of several lowest NFE states after electron doping for zigzag and armchair graphane nanoribbons, respectively. Here, we can see that the electron doping response behavior is difference between graphane and boron nitride nanoribbon superlattices. With electron doped, NFE-vacuum states in graphane nanoribbons can be occupied along with NFE-ribbon states. The reason why NFE-vacuum and NFE-ribbon states are not mixed as in boron nitride nanoribbons may because that graphane are thicker than boron nitride sheet, and the overlap between NFE-vacuum and NFE-ribbon states are smaller and their mixing thus become unfavorable.

We note that the response behavior observed here is also different from graphene nanoribbons, where, upon electron doping, the several lowest NFE states are all NFE-vacuum states. Here, the several lowest NFE-vacuum and NFE-ribbon states are very close in energy, although they are not mixed as in the boron nitride nanoribbon case. Both NFE-vacuum states and NFE-ribbon states become occupied upon electron doping. Therefore, the doped electron distribute both in vacuum and on ribbons. This is consistent with the change of electrostatic potential for graphane nanoribbons after electron doping. As shown in Fig.3 (a) and Fig.3 (b), with electron doping, the whole potential increases toward the vacuum level. While the potential in vacuum region and in the ribbon region increase almost at the same speed, and the shape of potential almost not changed after electron doping.

**CONCLUSION**

In conclusion, by first-principles calculation, we have identified two types of NFE states in graphane nanoribbon superlattices, and both are lowered upon electron doping. It is possible to use occupied NFE-vacuum states in graphane nanoribbon superlattices as a transport channel.

**ACKNOWLEGMENTS**

**FIGURE CAPTIONS**

FIG.1. (a) the geometric structure N=17 armchair graphane nanoribbon. (b) the band structure and Γ point density profiles of the five lowest NFE-ribbon states (lower panels) and the three lowest NFE-vacuum states (top panels) of N=17 armchair graphane nanoribbon supperlattices. (c) the geometric structure N=10 zigzag graphane nanoribbon. (d) the band structure and Γ point density profiles of a few lowest NFE states of N=10 zigzag graphane nanoribbon superlattice. The red and blue lines in the band structures represent NFE-ribbon and NFE-vacuum states, respectively. The gray dot-lines represent normal σ and π states.

FIG.2. (a) The band structures and the Γ point density profiles of a few lowest NFE states for (a) N=17 armchair graphane nanoribbon superlattice and (b) N=10 zigzag graphane nanoribbon superlattice with electron doping level at 0.02 electrons per carbon atom. The red and blue lines in the band structures represent the NFE-ribbon and NFE-vacuum states, respectively.

FIG.3. (a) The x-z plane averaged electrostatic potential for a unit cell of the N=17 armchair graphane nanoribbon superlattice and (b) with electron doping level of 0.02 electrons per carbon atom. The Fermi level is set to zero.

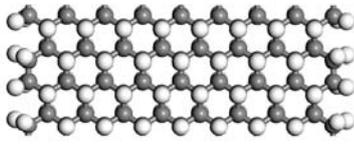
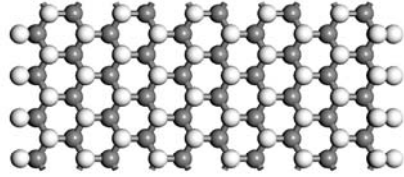
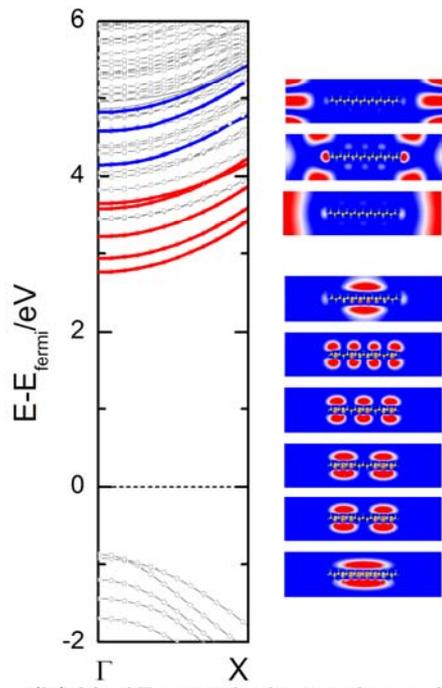
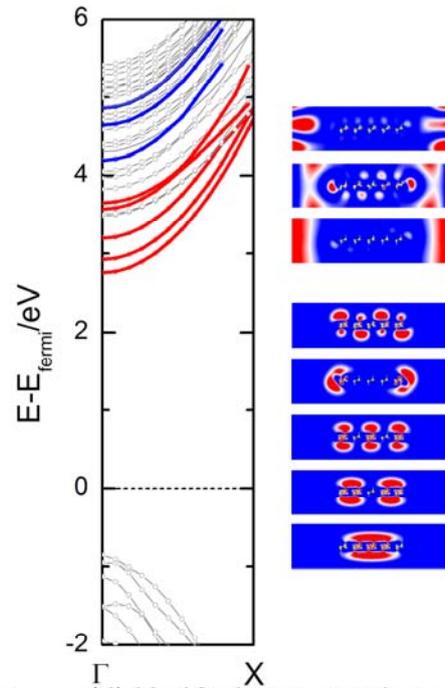

(b) N=17 armchair graphane NRs

(d) N=10 zigzag graphane NRs

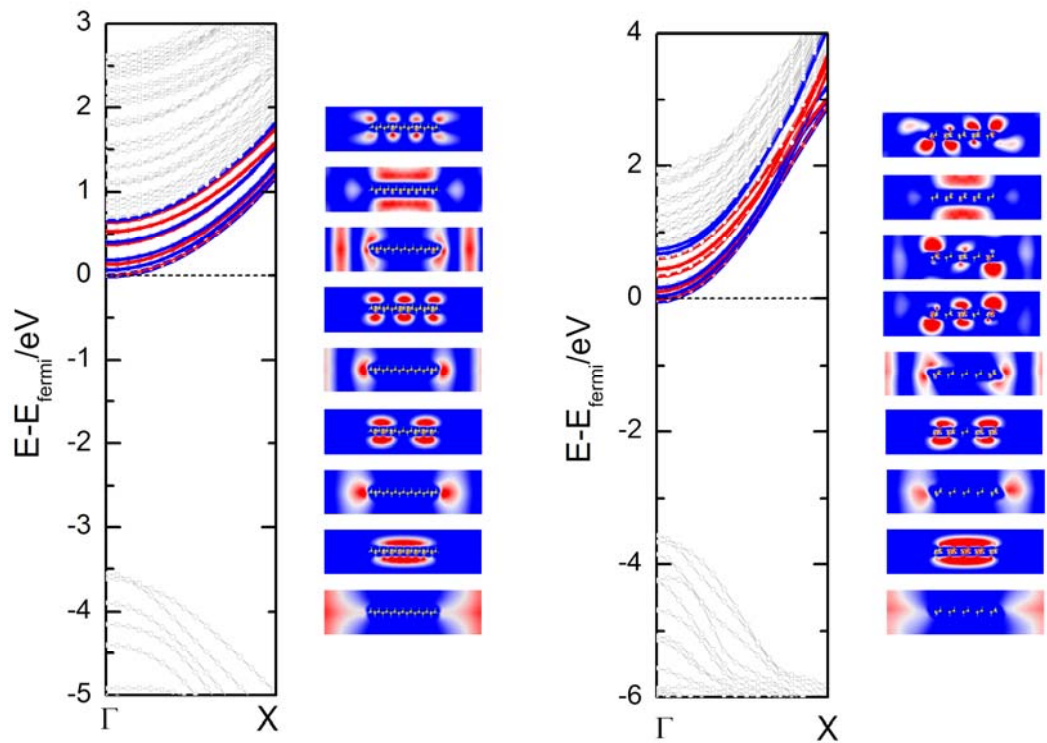

(a) N=17 armchair graphane NRs 0.02e/C  (b) N=10 zigzag graphane NRs 0.02e/C

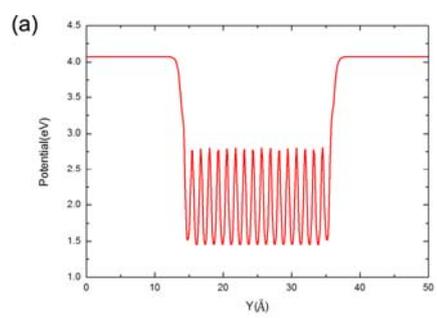 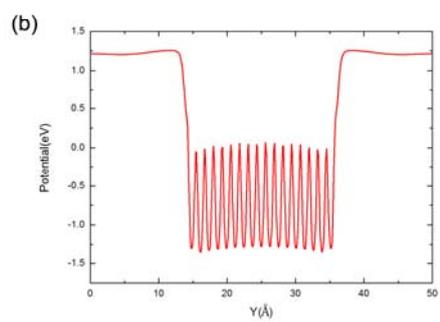